\documentclass[11pt]{article}

\usepackage[utf8]{inputenc}
\usepackage[T1]{fontenc}
\usepackage{amsmath,amssymb,amsthm}
\usepackage{graphicx}
\usepackage{booktabs}
\usepackage{multirow}
\usepackage{float}
\usepackage[hypertexnames=false]{hyperref}
\usepackage{url}
\usepackage{xspace}
\usepackage{microtype}
\usepackage{xcolor}
\usepackage[margin=1in]{geometry}
\usepackage{listings}

\definecolor{codebg}{gray}{0.97}
\definecolor{codeframe}{gray}{0.80}
\definecolor{codekeyword}{HTML}{2A4D6E}  
\definecolor{codestring}{HTML}{6B6B6B}   
\definecolor{codecomment}{HTML}{8A8A8A}  
\definecolor{codeprompt}{HTML}{555555}   

\lstdefinestyle{paperbase}{%
  basicstyle=\ttfamily\footnotesize,
  backgroundcolor=\color{codebg},
  frame=single,
  framerule=0.4pt,
  rulecolor=\color{codeframe},
  framesep=4pt,
  xleftmargin=6pt,
  xrightmargin=6pt,
  aboveskip=0.6\baselineskip,
  belowskip=0.6\baselineskip,
  breaklines=true,
  breakatwhitespace=false,
  breakindent=2em,
  prebreak=\mbox{\textcolor{codeframe}{$\hookleftarrow$}},
  columns=fullflexible,
  keepspaces=true,
  showstringspaces=false,
  upquote=true,
  tabsize=2,
  literate={→}{{$\rightarrow$}}{1}
           {←}{{$\leftarrow$}}{1}
           {≤}{{$\leq$}}{1}
           {≥}{{$\geq$}}{1}
           {×}{{$\times$}}{1}
           {…}{{\ldots}}{1},
}

\lstdefinestyle{shell}{%
  style=paperbase,
  language={},
  commentstyle=\color{codecomment}\itshape,
  morecomment=[l]{\#},
}

\lstdefinestyle{python}{%
  style=paperbase,
  language=Python,
  keywordstyle=\color{codekeyword},
  stringstyle=\color{codestring},
  commentstyle=\color{codecomment}\itshape,
}

\hypersetup{
    colorlinks=true,
    linkcolor=blue,
    filecolor=magenta,
    urlcolor=cyan,
    citecolor=blue,
    pdftitle={MimeLens: Position-Agnostic Content-Type Detection for Binary Fragments},
    pdfauthor={Michael J.\ Bommarito II},
    pdfsubject={Binary Analysis, Tokenization, BERT, MIME Classification, Benchmarking},
    pdfkeywords={binary analysis, byte-level pretraining, byte pair encoding, MIME classification, magika, libmagic, small encoders}
}

\newcommand{\byte}{\textsc{byte}\xspace}
\newcommand{\bpefour}{\textsc{bpe-4k}\xspace}
\newcommand{\bpesixteen}{\textsc{bpe-16k}\xspace}
\newcommand{\bpesixtyfour}{\textsc{bpe-64k}\xspace}
\newcommand{\tinysz}{\textsc{tiny}\xspace}
\newcommand{\smallsz}{\textsc{small}\xspace}
\newcommand{\mediumsz}{\textsc{medium}\xspace}
\newcommand{\mimebench}{MIME-125\xspace}

\newcommand{\magika}{\textsc{Magika}\xspace}
\newcommand{\libmagic}{\texttt{libmagic}\xspace}

\newcommand{\clspool}{\texttt{cls\_pool}\xspace}
\newcommand{\mimelens}{\textsc{MimeLens}\xspace}
\newcommand{\bmono}[1]{\nolinkurl{#1}}
\newcommand{\hfhub}{Hugging Face}

\title{\mimelens:\\
Position-Agnostic Content-Type Detection for Binary Fragments}

\author{
Michael J.\ Bommarito II\thanks{Portions of this work were prepared with assistance from large language models. The author is solely responsible for all content, including any errors or omissions.} \\
\texttt{michael.bommarito@gmail.com}
}

\date{May 2026}

\begin{document}

\maketitle

\begin{abstract}
File-type classification underlies many workflows like malware triage, forensic carving, packet inspection, and storage indexing. Learned systems such as Google's \magika{} assume whole-file access at a known offset, so they break on the inputs many of these tasks actually produce, like a single packet payload, a header-less carved fragment, a random disk block, or a chunked upload. We introduce \mimelens, a family of small BERT-style encoders pretrained on binary content from windows sampled at a uniformly random offset within each file, with no privileged head-of-file position, in standard- and short-context variants. A byte chunk goes in from anywhere in a file, no header needed and no fixed size; out comes one of \libmagic's $125$ MIME labels. On the clean head of complete files, \mimelens{} beats \magika{} v1.1 by $+10.7$~pp top-1 on \libmagic-labeled data, and it keeps classifying where \magika{} cannot: from a single mid-stream UDP packet, and more than twice as accurately as \libmagic{} and \magika{} on random mid-file disk blocks. The cost is latency: \mimelens{} runs roughly one to two orders of magnitude slower per sample on CPU than \magika{}, though it matches on consumer GPUs or in batch. All trained checkpoints are released on Hugging Face (\bmono{mjbommar/mimelens-001-*}).
\end{abstract}

\section{Introduction}
\label{sec:intro}

Content-type classification is the first and most foundational operation across domains as varied as reverse engineering and malware analysis, e-discovery, forensic carving, packet inspection, and storage optimization. The most canonical approach, \libmagic{}~\cite{libmagic}, which powers \texttt{file(1)} (Apache Tika~\cite{tika} takes the same magic-byte approach independently), uses hand-engineered byte-pattern rules anchored to fixed file offsets, organized around a 125-class MIME taxonomy. Google's recent \magika{} system~\cite{fratantonio2024magika} replaces \libmagic's rules with a $\sim 1$~M-parameter dense neural classifier over three 512-byte windows sampled from the file's head, middle, and tail, and reports $99\%$~F1 on Google's internal benchmark. \textbf{Both systems require whole-file access at a known file boundary.} A large class of real workloads yields only a chunk: packet payloads inspected mid-stream, forensic fragments recovered without a header, random disk-block reads at unknown offsets, streaming uploads not yet complete, or obfuscated chunks in a malware artifact, to name just a few.

This paper introduces \mimelens, a family of small ($\leq 38$~M-backbone-parameter) BERT-style encoders that classify file content type from a byte chunk taken at any offset, at the resolution of \libmagic's 125-class MIME taxonomy. We present results from \emph{three cells} selected from the broader family we trained: \mimelens-\mediumsz/\bpesixteen{} for clean-input classification, \mimelens-\mediumsz/\byte{} for streaming / partial-data inputs, and \mimelens-\mediumsz/\bpesixtyfour{} for throughput-bounded indexing, header-corruption-prone inputs, and forensic disk-block triage. We use the simplest setup: a byte chunk (e.g., a 4~KB buffer in our benchmarks) passes through the encoder, with or without BPE tokenization depending on the model, and the encoder emits a mean-pooled embedding that a linear-probe classifier maps to one of \libmagic's 125 MIME labels. These three are drawn from a factorial grid of $3$ model sizes $\times$ $4$ input pipelines $\times$ $2$ seeds ($3$ at the medium size), which we call the \emph{cube}. Results for all \mediumsz{} cells are in Appendix~\ref{app:family} and on Hugging Face.

The defining design property of \mimelens{} is its approach to pretraining data. Unlike other file-type models, the encoder is trained on fragments sampled from a uniformly random offset within each file (e.g., 1024 tokens for the main family, 256 for the short-context cells; Appendix~\ref{app:short_seq}). There is no privileged ``head of file'' or ``end of file'' position. The model is exposed to magic-byte regions, embedded-section payloads, mid-archive deflate streams, packed-binary entry-point neighborhoods, and arbitrary slices of long text and data containers with equal probability, weighted only by content-class balance. The consequence at inference time is that the same checkpoint that classifies the head of a complete file works on a $1.4$~KB packet payload mid-stream, a header-less carved fragment, a $4$~KB block from a damaged disk image, a random seek into a $50$~GB database, or a chunked HTTP body received before the whole file is in hand. The input is whatever bytes are on hand, with no fixed size. To our knowledge, no other deployed file-type classifier is trained on position-arbitrary inputs at this label resolution.

On a symmetric 25\% held-out test split of a stratified $4{,}096$-file benchmark drawn from the \textsc{magic-files} corpus (a \libmagic-labeled multi-source file set; Appendix~\ref{app:corpus}), with $n=1{,}024$, \mimelens-\mediumsz/\bpesixteen{} exceeds \magika{} v1.1 on \libmagic-labeled data at every level of stringency: aligned top-1 (a curated 21-class taxonomy-equivalence map applied symmetrically) $0.829$ vs $0.722$ ($+10.7$~pp); strict $0.828$ vs $0.653$ ($+17.5$~pp, of which most is \magika's taxonomy-naming mismatch); top-level $0.927$ vs $0.840$ ($+8.7$~pp). The map is fit on this corpus, so the aligned gap is diagnostic, not a frozen benchmark (§~\ref{sec:magika}); what would persist under a \magika{} retrained on \libmagic{} labels is an open question. The per-class breakdown (§~\ref{sec:magika}) shows the aggregate gap is non-uniform. \mimelens{} wins source-code and structured-text classes by $25$ to $90$~pp; \magika{} wins media, image, and script-language classes by $70$ to $85$~pp. For content mixes dominated by structured text \mimelens{} is the right tool; for mixes dominated by media, \magika{} remains the right tool. Two real-world validations exercise deployment regimes \magika{} is not designed for. 500 files transmitted as UDP datagrams over loopback ($n=498$ matched-by-stream-id after capture), classified from a \texttt{tcpdump} pcap (§~\ref{sec:network}), give \mimelens-\byte{} $0.855$ top-1 (Wilson 95\% CI $[0.82, 0.88]$) from a single 1.4~KB first packet. 1{,}000 random 4~KB blocks sampled from an unmounted \texttt{ext4} disk image (§~\ref{sec:disk}) give \mimelens-\bpesixtyfour{} $0.266$ mid-file top-1 (file-level cluster bootstrap CI $[0.22, 0.32]$, $B=1{,}000$) vs $0.093$ \libmagic{} and $0.112$ \magika.

The deployment-by-regime guidance is in §~\ref{sec:deployment}; §§~\ref{sec:network}--\ref{sec:disk} report the application studies in full; §~\ref{sec:system} describes the encoder family the deployment cells come from; §~\ref{sec:conclusion} catalogs the limitations and lists the released artifacts. The model family, within-cube results, methodological notes, and the size-scaling study are in Appendix~\ref{app:family}; the per-class breakdowns in Appendix~\ref{app:per_class}; the training configuration and model access in Appendix~\ref{app:reproduction}.

\paragraph{Artifacts.} All $36$ trained checkpoints are released on the \hfhub{} family hub; the deployed cells ship a baked $125$-class classifier head (Appendix~\ref{app:reproduction}). A reference implementation of the training stack is released on GitHub (\bmono{mjbommar/mimelens-training}); the $33$~GB corpus and the per-experiment artifacts are not released, as the corpus mixes content of varied provenance (Appendix~\ref{app:reproduction}).

\section{Related work}
\label{sec:related}

\paragraph{File-type identification.} \libmagic{}~\cite{libmagic}, the engine behind \texttt{file(1)}, is the de facto industrial reference for byte-pattern MIME identification; its rule database has accreted over three decades. Apache Tika~\cite{tika} takes the same magic-byte approach with an independent rule set. PRONOM and the corresponding DROID and Siegfried tools maintain an alternative format-identification registry oriented around digital-preservation use cases. TrID~\cite{trid} is a signature-based classifier widely used in forensic workflows; its label space is curated separately from \libmagic's. All four assume the input is a whole file with a known head offset. None is trained on position-arbitrary inputs at \libmagic's 125-class MIME granularity.

\paragraph{Learned file classifiers.} \magika{} v1~\cite{fratantonio2024magika} from Google is the closest published learned baseline: a $\sim 1$~M-parameter feedforward classifier over three 512-byte windows (head, middle, tail), trained on a Google-internal corpus. \magika's design assumes whole-file access; running it on a 4~KB chunk or a single packet payload collapses its head/middle/tail sampling onto three slices of \emph{that chunk}, not the whole file. We use \magika{} v1.1 as the primary calibration baseline (§~\ref{sec:magika}) and as a comparator in both application studies (§§~\ref{sec:network}--\ref{sec:disk}); it emits its own content-type taxonomy ($200{+}$ types), which we map onto \libmagic's labels for comparison, and the resulting numbers measure how \magika{} performs in regimes it was not designed for.

\paragraph{File-fragment classification.} A separate literature classifies fixed-size byte fragments without whole-file context, motivated by file carving and forensic recovery. Sceadan~\cite{beebe2013sceadan} uses concatenated $n$-gram features and an SVM at small label cardinality. Subsequent work has extended this with convolutional and recurrent neural baselines; the typical evaluation taxonomy is small (10--30 classes) and tied to the corpus of the cited paper rather than to a standardized MIME registry. Our position-arbitrary pretraining is in the spirit of this literature, but at \libmagic's much larger 125-class granularity and on a pretrained transformer backbone rather than a from-scratch supervised model.

\paragraph{Byte-level and BPE encoders.} Byte- and character-level tokenization-free pretraining (ByT5~\cite{xue2022byt5}, CANINE~\cite{clark2022canine}) and the byte-vs-BPE scaling trend (BLT~\cite{pagnoni2024blt}) motivate our input-pipeline axis. We consume the public binary-BPE tokenizers~\cite{bommarito2025binarybpe} directly and pretrain on a $33$~GB corpus anchored by the public \texttt{binary-30k} dataset~\cite{bommarito2025binary30k} ($29{,}793$ ELF/PE/Mach-O/APK binaries; the full training corpus, with packed binaries and drivers added, is not released). The byte-vs-BPE question itself is orthogonal to this paper's position-arbitrary deployment focus (Appendix~\ref{app:family}).

\section{The \mimelens{} encoder family}
\label{sec:system}

We call each (size, input pipeline, seed) combination a \emph{cell}.

\paragraph{Architecture.} All cells share a pre-norm transformer with RMSNorm without bias, rotary position embeddings (RoPE)~\cite{su2021rope} with $\theta = 10{,}000$, multi-head self-attention with head dimension fixed at $64$ across sizes, GeGLU feed-forward with expansion $8/3$, tied input/output embeddings, no biases, no dropout. The encoder produces a sequence of hidden states. We obtain a fixed-length representation by mean-pooling over body tokens (positions $1$ through $L{-}2$, excluding the CLS, SEP, and pad positions via the attention mask). We do not use the BERT-style \clspool{} linear projection (the dense layer applied to the CLS token): under MLM-only training it receives no gradient and stays at initialization (Appendix~\ref{app:methodological:clspool}).

\paragraph{Training recipe.} Every cell is pretrained MLM-only on the same $33$~GB multi-source binary corpus under matched compute ($22{,}888$ gradient steps at effective batch $128$ and sequence length $1024$ on a single consumer GPU), with only the tokenizer vocabulary and seed varying across cells. The full corpus composition, optimizer, masking schedule, and learning-rate schedule are specified in Appendix~\ref{app:reproduction}.

\begin{table}[ht]
\centering
\small
\setlength{\tabcolsep}{4pt}
\begin{tabular}{lrrrrl}
\toprule
cell & backbone & embed. & bytes/ & coverage & intended regime \\
     & params   & params & token  &          &                 \\
\midrule
\mediumsz/\byte{}         & 37.76M & 0.13M  & 1.000 & 1{,}022~B (25\%) & streaming / partial \\
\mediumsz/\bpesixteen{}   & 37.76M & 8.40M  & 1.727 & 1{,}765~B (43\%) & clean 4 KB head \\
\mediumsz/\bpesixtyfour{} & 37.76M & 33.56M & 2.087 & 2{,}134~B (52\%) & forensic / throughput \\
\bottomrule
\end{tabular}
\caption{The three deployed \mimelens{} cells (of $36$ released; Appendix~\ref{app:reproduction}). ``Backbone params'' excludes the embedding matrix; ``embedding params'' is the input-embedding matrix sized as (vocab~+~5~specials)~$\times d_\text{hidden}$. ``Bytes/token'' is the empirical average over the \textsc{magic-frags} corpus. ``Coverage'' is the bytes of a 4~KB buffer that the model's first 1{,}022 body tokens consume at seq\_len$=1024$. All three cells are seed~1 of the \mediumsz{} layer of the factorial cube (Appendix~\ref{app:family}).}
\label{tab:family}
\end{table}

The three cells share the same transformer body (37.76~M parameters, 12 layers, hidden $512$, 8 heads) and differ only in tokenizer vocabulary and embedding-matrix size: \byte{} uses raw \texttt{uint8} values in a 261-entry vocabulary; \bpesixteen{}/\bpesixtyfour{} use the public binary-tokenizer-001 BPE tokenizers~\cite{bommarito2025binarybpe}. Total params (backbone + embedding) therefore differ ($37.89$/$46.16$/$71.32$~M). ``Same-backbone-compute'' equalizes transformer-body FLOPs, not embedding-table size.

\paragraph{Effective byte coverage (caveat).} The 4~KB buffer is the same for all cells, but at the $1{,}022$-body-token limit each consumes a different prefix: \byte{} the first $\approx 1{,}022$~B ($25\%$), \bpesixteen{} $\approx 1{,}765$~B ($43\%$), \bpesixtyfour{} $\approx 2{,}134$~B ($52\%$). Byte-vs-BPE comparisons on the 4~KB benchmark are thus confounded by unequal coverage; a matched-$1{,}022$-byte-budget ablation is in Appendix~\ref{app:methodological:matched}.

\paragraph{Evaluation protocol.} All evaluations use \textsc{magic-files}, a \libmagic-labeled corpus of $28{,}498$ files across $105$ of \libmagic's $125$ MIME classes, with a stratified $4{,}096$-file headline benchmark and a $64$~KB-random-chunk variant \textsc{magic-frags} (composition in Appendix~\ref{app:corpus}). All \mimelens{} numbers in §§~\ref{sec:magika}--\ref{sec:disk} come from a \emph{frozen} encoder plus a \texttt{sklearn} logistic-regression (LR) probe on the mean-pooled embedding (\texttt{saga}, $75/25$ split, \texttt{random\_state}$=0$ for cube and Magika; contamination-clean LR-lbfgs for disk), labeled with the \libmagic-pinned 125-class taxonomy. End-to-end fine-tuning should be evaluated before deployment; we do not claim these frozen-probe numbers bound fine-tuned performance.

\section{Calibration against \magika}
\label{sec:magika}

We calibrate \mimelens-\mediumsz/\bpesixteen{} (the clean-input cell from §~\ref{sec:system}) against \magika{} v1.1~\cite{fratantonio2024magika} on \libmagic's 125-class MIME taxonomy. Both systems are evaluated on the same $25\%$ held-out split of the $4{,}096$-file \textsc{magic-files} corpus ($n=1{,}024$, same split as the cube's primary classification metric, §~\ref{sec:system}). We report top-1 under three matching rules, applied \emph{symmetrically} to both systems' predictions: \textbf{strict} (exact MIME string match), \textbf{aligned} (apply a curated 21-class equivalence map that collapses \magika~$\leftrightarrow$~\libmagic{} naming differences such as \texttt{text/x-python} $\equiv$ \texttt{text/x-script.python} and \texttt{application/x-dosexec} $\equiv$ \texttt{application/vnd.microsoft.portable-executable}), and \textbf{top-level} (compare only the top-level type: \texttt{text} vs \texttt{image} vs \texttt{application} vs $\ldots$).

\begin{table}[ht]
\centering
\small
\begin{tabular}{@{}lrrr@{}}
\toprule
matching rule & \mimelens & \magika & gap \\
\midrule
strict top-1                                          & \textbf{0.828} & 0.653 & $+17.5$~pp \\
aligned top-1 (21-class equiv. map, symmetric)        & \textbf{0.829} & 0.722 & $+10.7$~pp \\
top-level top-1 (text / image / application / \dots)  & \textbf{0.927} & 0.840 & $+8.7$~pp \\
\bottomrule
\end{tabular}
\caption{Apples-to-apples \mimelens-\mediumsz/\bpesixteen{} vs \magika{} v1.1 on the same $n=1{,}024$ held-out split. Both systems' predictions pass through the same equivalence map (``aligned''); the map raises \mimelens{} by $+0.1$~pp and \magika{} by $+7.0$~pp (the taxonomy-mismatch component). The $+10.7$~pp aligned gap is the residual after collapsing taxonomy-naming; since the map is fit on this corpus it is diagnostic, not a frozen benchmark number. \textbf{Bold: best per row.}}
\label{tab:magika-symmetric}
\end{table}

\paragraph{Interpretation.} \mimelens{} exceeds \magika{} at every stringency. \magika's $19.6$~pp strict-to-top-level lift ($0.653 \to 0.840$) is the taxonomy-mismatch component (it emits its own content-type names, e.g.\ \texttt{text/x-python} for \libmagic's \texttt{text/x-script.python}). The aligned $+10.7$~pp is the residual on this corpus after collapsing known mismatches; we do not claim it is what would persist under a \magika{} retrained on \libmagic{} labels. The $+8.7$~pp top-level lead shows the systems differ at coarse granularity too.

\paragraph{Status of the equivalence map.} The 21-class map was built from \magika's top-50 confusions on this evaluation corpus, not a held-out development split, so it is appropriate for decomposing the strict gap into taxonomy-mismatch and capability components but not as a publication-grade benchmark number; a different corpus may surface confusion pairs it does not cover. Future comparisons should freeze the map on a development split and report on the disjoint test split.

\paragraph{Per-class breakdown.} The aggregate $+10.7$~pp aligned gap is not uniformly distributed across MIME classes. Table~\ref{tab:magika-perclass} shows the 8 largest cross-system aligned-top-1 differences (4 per direction) on classes with $\geq 60$ samples in the full corpus.

\begin{table}[h]
\centering
\small
\setlength{\tabcolsep}{4pt}
\begin{tabular}{@{}lrrrr@{}}
\toprule
class & $n$ & \mimelens & \magika & gap \\
\midrule
\multicolumn{5}{@{}l}{\emph{Largest \mimelens{} wins}} \\
\texttt{text/x-file}             & 61  & 0.885 & 0.000 & $+0.885$ \\
\texttt{text/x-c}                & 122 & 0.680 & 0.041 & $+0.639$ \\
\texttt{text/x-c++}              & 90  & 0.578 & 0.000 & $+0.578$ \\
\texttt{application/vnd.ms-excel}& 138 & 0.971 & 0.572 & $+0.399$ \\
\midrule
\multicolumn{5}{@{}l}{\emph{Largest \magika{} wins}} \\
\texttt{text/x-perl}             & 100 & 0.150 & 1.000 & $-0.850$ \\
\texttt{image/webp}              & 169 & 0.166 & 1.000 & $-0.834$ \\
\texttt{text/x-php}              & 108 & 0.176 & 1.000 & $-0.824$ \\
\texttt{image/gif}               & 133 & 0.218 & 1.000 & $-0.782$ \\
\bottomrule
\end{tabular}
\caption{Per-class aligned top-1, classes with $\geq 60$ samples (8 largest differences, 4 per direction). Per-class numbers are computed via 5-fold cross-validation over the full 4{,}096-file corpus (out-of-fold predictions), \emph{not} the 75/25 held-out split used for the aggregate numbers above; the wider per-class denominators give stable enough per-class supports to be informative. \mimelens{} dominates source-code and office-document classes by $25$ to $90$~pp; \magika{} dominates media, image, and script-language classes by $70$ to $85$~pp.}
\label{tab:magika-perclass}
\end{table}

A practitioner whose disk content is mostly source code or office documents will see \mimelens{} dominate; a practitioner whose content is mostly media or web-scripting languages will see \magika{} dominate. The aggregate gap should not be read as uniform superiority.

\paragraph{Scope.} All ground truth here is \libmagic-derived, so these numbers measure how well \mimelens{} imitates a \libmagic-pinned taxonomy on a corpus labeled by that same pipeline, the right target for a \libmagic-based consumer (\texttt{file(1)}, indexing pipelines, malware-triage front ends). They are not a generalization claim against independent ground truth (PRONOM/Siegfried, DROID, IANA, human adjudication were not used; §~\ref{sec:conclusion}). Read them as ``\mimelens{} is a better \libmagic-style classifier than \magika{} on this set,'' not ``closer to objective file-type truth.''

\section{Applications}
\label{sec:applications}

The calibration in §~\ref{sec:magika} measured \mimelens{} on the clean head of complete files, the regime \magika{} was built for. We now turn to two regimes the whole-file classifiers were not built for: live network packet capture (§~\ref{sec:network}) and forensic disk-block reads (§~\ref{sec:disk}). Each classifies identical bytes with \mimelens{} and the \magika{}/\libmagic{} baselines.

\subsection{Network packet classification}
\label{sec:network}

Several real workloads classify file content from packets on the wire before the whole file is available: deep packet inspection in security middleboxes, content-aware routing in CDN edges, MIME-typing of streaming uploads, and forensic packet capture. The input is one or more MTU-sized payloads with a known transmit order but no guarantee that any single read covers a full file. This regime is what the position-arbitrary pretraining is meant to serve: a model that classifies any 4~KB byte window should classify the first $1{,}448$ bytes (one Ethernet MTU payload) of a file just as well as the first $4{,}096$.

\paragraph{Protocol.} 500 deterministically-sampled files from a held-out subset of the \textsc{magic-files} corpus, transmitted as UDP datagrams from \texttt{127.0.0.1} to \texttt{127.0.0.1:9999}; each datagram carries a 12-byte header (\texttt{\{stream\_id, seq\_no, total\_n\}}) followed by $\leq 1{,}448$~B of payload (a $1500$-byte IP MTU less $20$~B IP, $8$~B UDP, our $12$~B header, and a small safety margin). Files are capped at $100$~KB. \texttt{tcpdump -i lo -U -s 0 -w capture.pcap "udp port 9999"} runs alongside the sender. The resulting pcap ($\sim$26~MB, $17{,}510$ datagrams) covered $498$ of the $500$ sent streams cleanly; two streams had at least one packet dropped at the capture layer and were excluded from the analysis. The per-stream in-flight delivery itself was lossless and in-order (loopback). For each captured stream we sort packets by \texttt{seq\_no}, concatenate the first $K$ payloads, and truncate or right-zero-pad to $4{,}096$~bytes; the classifier sees this prefix buffer at thresholds $K \in \{1, 2, 3, 5, 10, \text{all}\}$.

\paragraph{Results.}

\begin{table}[ht]
\centering
\small
\setlength{\tabcolsep}{4pt}
\begin{tabular}{lrrrrrrr}
\toprule
$K$ & \byte & \bpefour & \bpesixteen & \bpesixtyfour & \magika & \libmagic & TrID (sc) \\
\midrule
$1$ (1.4 KB)  & \textbf{0.855} & 0.815 & 0.809 & 0.745 & 0.588 & 0.739 & 0.713 \\
              & {\scriptsize[.82,.88]} & {\scriptsize[.78,.85]} & {\scriptsize[.77,.84]} & {\scriptsize[.71,.78]} & {\scriptsize[.55,.63]} & {\scriptsize[.70,.78]} & {\scriptsize[.67,.75]} \\
$3$ (4.3 KB)  & \textbf{0.855} & 0.821 & 0.821 & 0.763 & 0.618 & 0.791 & 0.729 \\
all           & \textbf{0.855} & 0.821 & 0.821 & 0.763 & 0.618 & 0.791 & 0.729 \\
              & {\scriptsize[.82,.88]} & {\scriptsize[.79,.85]} & {\scriptsize[.79,.85]} & {\scriptsize[.73,.79]} & {\scriptsize[.58,.66]} & {\scriptsize[.75,.83]} & {\scriptsize[.69,.77]} \\
\bottomrule
\end{tabular}
\caption{Network-capture top-1 by system and $K$, $n=498$ streams. All seven systems classified on the same captured prefix bytes (4~KB buffer, right-zero-pad). The first six accuracy columns are top-1 against the \libmagic-pinned ground-truth MIME directly. The \texttt{TrID (sc)} column is \emph{self-consistency rate}, not accuracy: it counts streams for which TrID's prefix prediction agrees with TrID's full-stream prediction and is not \texttt{Unknown}. TrID's label space is incomparable to \libmagic-125, so a top-1-vs-truth metric is not available. Wilson 95\% CIs in brackets for $K=1$ and $K=\text{all}$. \textbf{Bold: best accuracy per row} (among the six accuracy columns; TrID sc shown separately).}
\label{tab:network}
\end{table}

\paragraph{Single-packet saturation.} \mimelens-\byte{} hits its asymptote ($0.855$) at $K=1$ and stays flat: the byte tokenizer is the identity, so the encoder consumes the first $1{,}022$ bytes that arrive, and a single $1{,}448$-byte datagram more than covers them. The byte cell is architecturally a $1{,}022$-byte classifier here. The BPE cells reach $\sim 1{,}766$ bytes into the buffer, so they climb from $K=1$ to $K=3$ as real bytes replace the zero-padding, then plateau once $K \times 1{,}448 \geq 4{,}096$.

\begin{figure}[h]
\centering
\includegraphics[width=0.85\textwidth]{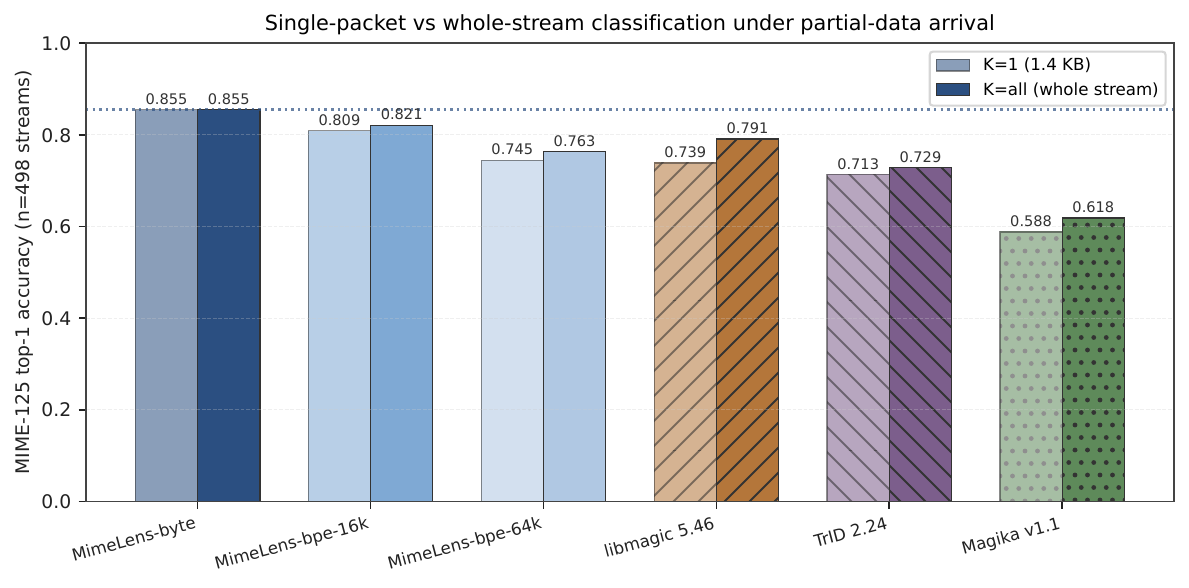}
\caption{Single-packet ($K=1$, light bars) vs whole-stream ($K=\text{all}$, dark bars) top-1 accuracy on real captured UDP traffic ($n=498$ streams), against the \libmagic-pinned ground truth. TrID is omitted here (its label space has no top-1-vs-truth metric; see the self-consistency column of Table~\ref{tab:network}). Dotted horizontal line marks \mimelens-\byte{} at $0.855$.}
\label{fig:network_bars}
\end{figure}

\paragraph{Scope.} This is an early-prefix cleartext UDP regime: in-order loopback delivery, no loss or reordering, $K=1$ always containing the header. It validates that \mimelens-\byte{} survives MTU-sized fragmentation, not arbitrary-offset packet inspection; read it as ``classifies the early prefix of a clean transfer from one packet,'' not ``classifies any packet from any flow.'' Random-offset sampling, TCP segmentation, loss/reordering, and encrypted transport are deferred (§~\ref{sec:conclusion}).

\subsection{Forensic disk-block classification}
\label{sec:disk}

Forensic-triage tools (The Sleuth Kit, Autopsy, EnCase) routinely classify content from random reads of raw disk blocks where no magic byte is at the read offset and file boundaries are unknown. \libmagic{} on a random $4$~KB mid-file block typically returns \bmono{application/octet-stream} (its default for unidentified content); \magika{} requires whole-file access and its head/middle/tail sampling strategy degenerates when given $4$~KB of unknown-offset bytes. Position-arbitrary pretraining targets exactly this case: a mid-file $4$~KB block carries no header and no offset cue, and the encoder was trained on precisely such windows.

\paragraph{Protocol.} We build a $1$~GB \texttt{ext4} image (no journal, $4{,}096$~B block size) populated with $3{,}066$ files from a held-out subset of the \textsc{magic-files} corpus. File selection caps any single file at $10$~MB and any single MIME at $5\%$ of the byte budget, producing a roughly balanced corpus across $99$ MIME classes ($\leq 5\%$ each). We unmount, then sample $1{,}000$ random $4$~KB block offsets uniformly across the byte range, skipping the boot/super-block region and all-zero blocks; on the headline \texttt{ext4}/sequential-bulk cell $995$ of $1{,}000$ blocks fall in allocated regions ($980$ of them mid-file, $15$ at the start of an owning file) and the remaining $5$ are unallocated.

Each block is classified by all four systems on identical bytes. Ground truth is the owning file's MIME from the \libmagic{} label set, obtained by parsing the unmounted image with The Sleuth Kit's \texttt{fls} + \texttt{istat}. The identical build--unmount--sample--classify--label pipeline is repeated on a matching $1$~GB \texttt{NTFS} image populated from the same held-out corpus. The Sleuth Kit's \texttt{fls} + \texttt{istat} parse the NTFS \texttt{\$MFT} the same way they parse \texttt{ext4} inodes, so the block-to-owner mapping and ground-truth labeling are method-identical across filesystems. The full filesystem~$\times$~init-strategy matrix (\texttt{ext4} and \texttt{NTFS}) is reported in the consistency paragraph below.

The logistic-regression probe is re-fit per cell with every \texttt{sha256} that appears on the disk image excluded from the probe-training pool, so the head bytes of on-disk files leak nothing into the classifier. Confidence intervals are file-level cluster bootstrap ($B=1{,}000$, resample by owning-file), which is the statistically correct unit when multiple sampled blocks come from the same source file: the $980$ mid-file blocks of the headline cell come from $254$ unique files (mean $3.9$ blocks per file, max $36$).

\paragraph{Headline result (\texttt{ext4} / sequential-bulk).}

\begin{table}[ht]
\centering
\small
\begin{tabular}{@{}lrr@{}}
\toprule
system & top-1 & 95\% cluster CI \\
\midrule
\mediumsz/\bpesixtyfour{} & \textbf{0.266} & $[0.22, 0.32]$ \\
\mediumsz/\bpesixteen{}   & 0.220 & $[0.17, 0.27]$ \\
\mediumsz/\byte{}         & 0.219 & $[0.17, 0.27]$ \\
\libmagic{} 5.46          & 0.093 & $[0.06, 0.13]$ \\
\magika{} v1.1            & 0.112 & $[0.08, 0.16]$ \\
\bottomrule
\end{tabular}
\caption{Per-block top-1 MIME-125 on the $980$ mid-file blocks of a $1$~GB \texttt{ext4} image. Per-cell contamination-clean probe; 95\% file-level cluster bootstrap CIs in brackets ($B=1{,}000$). All three \mimelens-\mediumsz{} cells exceed both comparators with non-overlapping CIs. Within \mimelens{}, \bpesixtyfour{} point estimate leads but file-level CIs overlap with \byte{}.}
\label{tab:disk-headline}
\end{table}

\begin{figure}[h]
\centering
\includegraphics[width=0.95\textwidth]{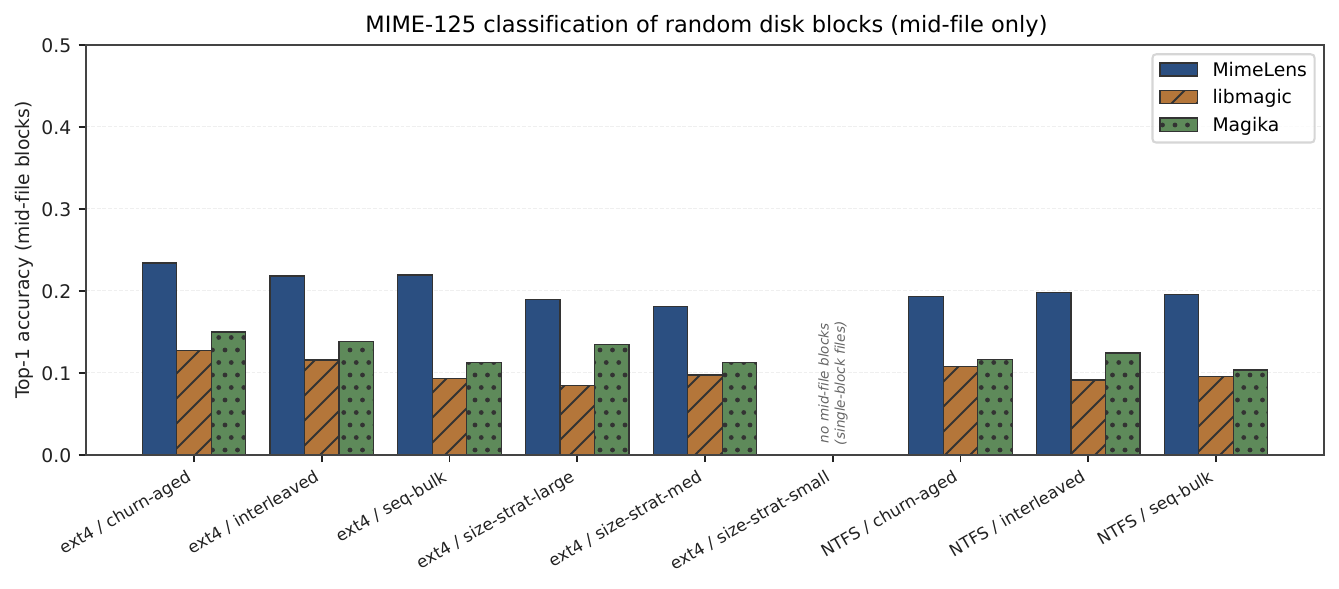}
\caption{Mid-file top-1 across the nine-cell disk-fingerprint matrix (\texttt{ext4} cells under four init-strategies plus two size-stratified sub-cells; \texttt{NTFS} cells under three init-strategies). \mimelens-\byte{} (blue) is at $0.18$--$0.23$ across the eight cells with mid-file data (the \texttt{ext4}/size-stratified-small cell has single-block files only, so no mid-file blocks, and is shown empty); \libmagic{} (orange) and \magika{} (green) sit at $0.09$--$0.15$. Error bars are file-level cluster $95\%$ CIs; \mimelens{} bars do not overlap with the comparator bars on any cell.}
\label{fig:disk_matrix}
\end{figure}

\paragraph{Matrix-cell consistency.} The headline result replicates across nine matrix cells: \texttt{ext4} run with each of the four init-strategies (sequential-bulk, interleaved-fragmented, churn-aged, size-stratified) plus two extra ext4 size-stratified sub-cells (small-files and large-files variants), and \texttt{NTFS} run with three of the four init-strategies (sequential-bulk, interleaved-fragmented, churn-aged; the NTFS resident-attribute path is incompatible with size-stratified and was skipped). Each cell samples $1{,}000$ blocks from the same MIME-balanced corpus (the mid-file subset varies by cell). Figure~\ref{fig:disk_matrix} shows \mimelens-\byte{}, \libmagic{}, and \magika{} mid-file top-1 with cluster-bootstrap error bars. The \byte{} cell stays in $0.18$--$0.23$ across the eight matrix cells with mid-file data, and never overlaps with the comparators ($0.09$--$0.15$).

\paragraph{Per-class structural shape.} The \byte{} cell's aggregate $\sim 0.22$ hides large per-class heterogeneity (per-class table in Appendix~\ref{app:per_class:disk}): \mimelens-\byte{} reaches $\geq 0.78$ on text-structured classes (\texttt{image/svg+xml} $0.98$, \texttt{text/csv} $0.88$, \texttt{text/plain} $0.84$, \bmono{application/octet-stream} $0.79$) and hits $0$ on classes whose mid-file content is high-entropy noise (\texttt{zip}, \texttt{7z}, \texttt{encrypted}, \texttt{jpeg}, \texttt{flac}). The aggregate is the weighted average of ``near-perfect on text-like'' and ``zero at high-entropy.'' A disk of mostly source/HTML/CSV/office-XML exceeds the headline; a disk of mostly media floors across all systems.

\paragraph{LUKS-encrypted negative control.} On $1{,}000$ random blocks of ciphertext from a LUKS-encrypted \texttt{ext4} volume (same corpus, raw-image read), \mimelens-\byte{} returns \bmono{application/octet-stream} for $99.3\%$ and a compressed-archive MIME for the rest; prediction-distribution entropy is $0.07$~bits, against a $6.70$-bit maximum over $99$ classes, mean confidence $0.935$. The model collapses onto an ``opaque'' surrogate with high confidence rather than a calibrated unknown, which is acceptable only if the pipeline routes \bmono{application/octet-stream} and the high-entropy archive labels to an abstention stage. We saw no class-prior bias on random bytes that would be a fingerprinting concern at this $n$.

\paragraph{Operational interpretation.} These embeddings call for a three-stage classifier, not a single top-1 system: (1)~opaque-detection, abstaining on blocks whose entropy looks compressed/encrypted/media; (2)~classify-when-confident above a per-class threshold (rejecting the bottom-$20\%$ confidence band lifts accepted-set top-1 by $5$--$10$~pp in the cube calibration sweep); (3)~an explicit ``opaque'' label for the rest. That yields a smaller but trustworthy ``confident MIME'' set plus a larger ``opaque'' set, which is what a carving pipeline wants.

\paragraph{Scope.} This is a forensic-stress workload, not a typical disk: a $5\%$-of-bytes cap per MIME class (an unconstrained image let one $700$~MB ZIP take $88\%$ of bytes) over a corpus that over-represents compressed and encrypted formats. Absolute mid-file accuracy ($\sim 0.22$--$0.27$ depending on cell, over $99$ MIME classes) is far below the clean-input numbers (§~\ref{sec:magika}) because mid-file blocks of compressed and encrypted formats are information-theoretically opaque to any classifier. The relative win is the only useful comparison in this regime.

\section{Deployment guidance}
\label{sec:deployment}

The calibration (§~\ref{sec:magika}) and the two application studies (§§~\ref{sec:network}--\ref{sec:disk}) each point to a different cell. The full model family (all $36$ checkpoints) is released on \hfhub{} (Appendix~\ref{app:reproduction}); which one to deploy depends on the input regime, and three \mediumsz{} cells cover the common cases. For clean, complete inputs classified from the head of a file, \mediumsz/\bpesixteen{} is the most accurate choice: it beats \magika{} v1.1 by $+10.7$~pp aligned top-1 on \libmagic-labeled data and wins at every level of stringency (§~\ref{sec:magika}), and it ties the \byte{} cell once byte coverage is matched. When data arrives a packet or a partial upload at a time, \mediumsz/\byte{} is preferable: because its tokenizer is the identity map, the cell is effectively a $1{,}022$-byte classifier, so a single $1.4$~KB datagram already saturates its input and yields $0.855$ top-1 from one packet (§~\ref{sec:network}). For throughput-bounded indexing at petabyte scale, \mediumsz/\bpesixtyfour{} packs roughly twice as many bytes into each token as the byte cell, trading about $6$~pp of clean-input accuracy for proportionally fewer forward passes.

The \bpesixtyfour{} cell is also the best choice for the harder, position-arbitrary regimes. On random mid-file disk blocks it roughly doubles both \libmagic{} and \magika{} with non-overlapping confidence intervals (§~\ref{sec:disk}); its absolute mid-file accuracy is only $0.27$, so it is best deployed behind a confidence-thresholded abstain stage rather than as a raw top-1 classifier. It is likewise the most robust cell to header corruption, losing $7$--$14$~pp under header zeroing where the byte cell loses up to $30$~pp (Table~\ref{tab:short_seq_adv}), because its wider tokens carry signal from deeper into the window. For sub-KB inputs such as DNS payloads, sub-MTU packets, or small carve fragments, the seq\_len$=256$ short-context cells match or exceed the parent cube at a quarter of the context and roughly $4$--$5\times$ lower CPU latency (Appendix~\ref{app:short_seq}). \mimelens{} is not always the right answer: for sub-millisecond whole-file triage over broad categories \magika{} remains preferable ($\sim 1.3$~ms versus $202$~ms per sample on CPU), and a deployment dominated by media, image, or web-scripting content will see \magika{} win on per-class accuracy as well (§~\ref{sec:magika}). The encoder closes that latency gap only on a GPU or in batch.

All three deployed cells are \mediumsz, the largest size in the cube, and the size sweep in Appendix~\ref{app:size_regime} shows this is not uniform overkill. Single-packet classification scales cleanly with backbone size, from $0.685$ at \tinysz{} through $0.739$ at \smallsz{} to $0.855$ at \mediumsz{} with non-overlapping Wilson intervals, so the largest backbone earns its place there. Random disk blocks tell the opposite story: the three sizes land between $0.23$ and $0.27$ and are statistically indistinguishable, each still beating both baselines by at least $2\times$, so a throughput-bound disk-triage deployment can fall back to the $3.15$~M \tinysz{} cell, twelve times smaller, at no measurable cost.

Two caveats bound these recommendations. Every number reported here comes from a frozen encoder and a logistic-regression probe, so end-to-end fine-tuning on the production label distribution is the recommended pre-deployment step and may shift the figures. And because all ground truth is \libmagic-derived, the comparisons measure agreement with a \libmagic-style taxonomy rather than objective file-type truth; the next section lists the full limitations.

\section{Conclusion}
\label{sec:conclusion}

\mimelens{} is a family of small BERT-style encoders that classify file content type from a byte chunk taken at any offset, at \libmagic's 125-class MIME resolution. Across the calibration (§~\ref{sec:magika}) and the two application studies (§§~\ref{sec:network}--\ref{sec:disk}), one design choice carries the paper: position-arbitrary pretraining. By training on windows drawn from uniformly random offsets within each file, with no privileged head-of-file position, the encoder learns to label a chunk wherever it comes from. This is what separates it from whole-file classifiers such as \magika{} and \libmagic, and what lets it serve the streaming, fragment, packet-payload, random-seek, and header-corrupted regimes those systems were not built for. On clean inputs it is competitive with the state of the art, exceeding \magika{} on \libmagic-labeled data at every level of stringency on the 4~KB head (§~\ref{sec:magika}); on inputs the whole-file systems cannot handle it is well ahead, classifying a file from a single $1.4$~KB packet (§~\ref{sec:network}) and roughly doubling both baselines on random mid-file disk blocks (§~\ref{sec:disk}). A short-context variant trained at seq\_len$=256$ extends the same behavior to sub-KB inputs at several times lower latency (Appendix~\ref{app:short_seq}).

These results come with clear boundaries. All ground truth is \libmagic-derived, so the comparisons establish that \mimelens{} is a better \libmagic-style classifier than \magika{} on this corpus, not that it is closer to objective file-type truth; cross-validation against PRONOM, Siegfried, or human-adjudicated labels remains future work. The evaluation splits are disjoint by SHA-256 but not by source family, which may inflate accuracy on text-structured classes that share boilerplate. The two application studies are deliberately narrow: the network study is early-prefix cleartext UDP over loopback, and the disk study uses a forensic-stress distribution that over-represents compressed and encrypted formats, so arbitrary-offset packets, TCP segmentation, loss and reordering, encrypted transport, and realistic disk mixes are all left open. Every reported figure is a frozen-probe number, and end-to-end fine-tuning would move it. On CPU the encoder is roughly one to two orders of magnitude slower per sample than \magika, a gap that closes only on a GPU or in batch, so \mimelens{} is the wrong tool for sub-millisecond whole-file triage.

All $36$ trained checkpoints are publicly released on the Hugging Face family hub \bmono{mjbommar/mimelens-001-*}, spanning the three model sizes, four input pipelines, and both context lengths; the deployed cells also ship a baked $125$-class classifier head, and every cell exposes the mean-pooled encoder for custom probing or fine-tuning (Appendix~\ref{app:reproduction}). A reference implementation of the training stack is released on GitHub (\bmono{mjbommar/mimelens-training}); the $33$~GB corpus is not. \mimelens{} does not aim to replace \magika{} or \libmagic; it occupies a different point on the latency, granularity, and position-arbitrariness surface, and is built for the growing set of workloads that must classify content from a fragment rather than a whole file.

\bibliographystyle{plain}
\bibliography{bibtex/references}

\clearpage
\appendix
\section{Model family and within-cube results}
\label{app:family}

The three deployed cells are seed-1 of the \mediumsz{} layer of a $3 \times 4 \times \{2,3\}$ factorial cube: three model sizes (\tinysz{} $3.15$~M backbone / $4$ layers; \smallsz{} $14.16$~M / $8$; \mediumsz{} $37.76$~M / $12$, head dim $64$, hidden $512$) crossed with four input pipelines (raw bytes; binary BPE at $4{,}101$, $16{,}391$, $65{,}543$ vocabulary including $5$ special tokens) and two pretraining seeds, extended to three at \mediumsz. Twenty-eight checkpoints (the parent cube; the eight short-sequence cells of Appendix~\ref{app:short_seq} bring the released total to $36$), each trained for $22{,}888$ gradient steps at effective batch $128$, sequence length $1024$, bf16, with architecture, optimizer, masking, and the $33$~GB corpus held constant. Matched-compute is along the gradient-step axis; BPE cells see $\sim 2\times$ more byte content per step, a confound the matched-coverage note below addresses.

\begin{table}[h]
\centering
\small
\begin{tabular}{lrrrr}
\toprule
size & \byte & \bpefour & \bpesixteen & \bpesixtyfour \\
\midrule
\tinysz   & 0.740 & 0.732 & \textbf{0.747} & 0.724 \\
\smallsz  & 0.771 & 0.757 & 0.751 & \textbf{0.792} \\
\mediumsz & 0.799 & 0.783 & \textbf{0.808} & 0.738 \\
\bottomrule
\end{tabular}
\caption{\mimebench head top-1 (LR-saga, 75/25 split, \texttt{random\_state}=0) per cell, seed-mean. \mediumsz/\bpesixteen{} (general classification) and \mediumsz/\byte{} (streaming) are deployed; \mediumsz/\bpesixtyfour{} regresses on this metric but is deployed for throughput, header-robustness, and disk-block triage.}
\label{tab:family_top1}
\end{table}

\subsection{Within-cube findings}
Classification top-1 (Table~\ref{tab:family_top1}) rises modestly with size, $+6$ to $+7$~pp from \tinysz{} to \mediumsz{} on the best vocabulary, with diminishing returns. \bpesixteen{} or \byte{} lead at \mediumsz; \bpesixtyfour{} wins at \smallsz{} but gives back $\sim 5$~pp from \smallsz{} to \mediumsz. The byte-vs-BPE retrieval and language-modeling metrics are orthogonal to deployment and omitted here.

\subsection{Methodological notes}
\label{app:methodological}
\paragraph{No \clspool{} gradient.}\label{app:methodological:clspool} Under MLM-only training the BERT-style \clspool{} projection receives no gradient and stays at initialization, $\|W^{\text{trained}}_{\text{cls\_pool}} - W^{\text{init}}_{\text{cls\_pool}}\|_\infty = 0$ across all $28$ cells; we mean-pool body tokens instead.

\paragraph{Matched byte coverage.}\label{app:methodological:matched} The headline $4$~KB benchmark gives BPE cells $1.7$--$2.1\times$ more effective byte coverage than \byte{} (§~\ref{sec:system}). Capping every cell to a common $1{,}022$-byte budget, \byte{} and \bpesixteen{} become statistically tied, so the clean-head \bpesixteen{} edge is largely a coverage effect, not a representational one.

\subsection{Inference latency}
Single-sample latency for \mediumsz/\bpesixteen{} on an idle Intel i9-12900K (torch 2.11.0, onnxruntime 1.26.0) is $202$~ms (PyTorch fp32) and $382$~ms (ONNX dynamic int8) at seq\_len$=1024$, versus \magika{} v1.1 at $\sim 1.3$~ms/sample on the same CPU ($\sim 155\times$). Dynamic int8 is \emph{slower} than fp32 at seq\_len$=1024$ on both the i9-12900K and an AMD Ryzen~7~7840HS (two AVX-VNNI CPUs), so the overhead is intrinsic to dynamic quantization at this context length, not a missing-VNNI effect; at seq\_len$=256$ int8 is faster ($26$ vs $46$~ms). The CPU cost is a CPU artifact: on an NVIDIA RTX~4060~Ti the same encoder runs at $8.7$~ms single-sample (seq\_len$=1024$) and $1.5$~ms/sample batched (seq\_len$=256$, batch~$64$), within range of \magika's CPU speed. \mimelens{} trades CPU latency for a $125$-class taxonomy and position-arbitrary inputs (§~\ref{sec:deployment}); on GPU, or batched, the trade narrows sharply.

\subsection{Short-sequence (seq\_len$=256$) cells}
\label{app:short_seq}
We retrained the \mediumsz{} tier at seq\_len$=256$ ($4$ vocabularies $\times\,2$ seeds, $8$ released cells, \texttt{-seq256} suffix) for the sub-KB regimes where the parent cube's $1{,}022$--$2{,}134$ byte coverage is excess: DNS payloads, sub-MTU packets, small carve fragments. On the held-out $4$~KB head (Table~\ref{tab:short_seq_adv}) the $256$-token BPE cells tie or beat the $1024$-token parent, \bpesixtyfour{} by $5.5$~pp, and the byte cell ties. From a single $1.4$~KB UDP packet (Table~\ref{tab:short_seq_net}) the $256$-token cells reach $0.76$--$0.79$, and \bpesixtyfour{} is the only cell whose packet accuracy improves as context shrinks. CPU latency drops $\sim 4$--$5\times$ at seq\_len$=256$ (CPU: fp32 $202 \to 46$~ms; int8 $382 \to 26$~ms, now faster than fp32), so the $256$-token int8 cell is $\sim 20\times$ \magika{} per sample versus $\sim 155\times$ for the parent cube.

\begin{table}[h]
\centering
\small
\setlength{\tabcolsep}{4pt}
\begin{tabular}{@{}lrrrrrrrr@{}}
\toprule
cell @ s1 & 1024 clean & 256 clean & $\Delta$clean & 256 $\Delta_\text{z4}$ & $\Delta_\text{z16}$ & $\Delta_\text{z64}$ & $\Delta_\text{r4}$ & $\Delta_\text{8K-prepend}$ \\
\midrule
\byte         & $0.813$ & $0.817$ & $+0.4$ & $-3.5$ & $-9.2$ & $\mathbf{-29.7}$ & $-2.7$ & $-78.7$ \\
\bpefour      & $0.793$ & $0.810$ & $\mathbf{+1.7}$ & $-3.7$ & $-6.0$ & $-12.4$ & $-1.6$ & $-80.5$ \\
\bpesixteen   & $0.799$ & $0.800$ & $+0.1$ & $-1.4$ & $-6.0$ & $-12.6$ & $-1.5$ & $-79.5$ \\
\bpesixtyfour & $0.727$ & $0.782$ & $\mathbf{+5.5}$ & $-2.9$ & $-7.2$ & $-14.4$ & $-2.5$ & $-77.7$ \\
\bottomrule
\end{tabular}
\caption{Held-out top-1 + adversarial perturbations on the $4$~KB head benchmark, $n=4{,}096$ files, $75/25$ stratified split (\texttt{random\_state}=0). Numbers under perturbation variants are $\Delta$top-1 from the clean s1 number (\texttt{z4}/\texttt{z16}/\texttt{z64} zero the first $4$/$16$/$64$ input bytes; \texttt{r4} randomizes the first $4$; \texttt{8K-prepend} prepends $8$~KB of unrelated bytes). 1024-cube clean numbers come from Table~\ref{tab:family_top1}. Bold $\Delta$~clean = seq\_len$=256$ beats seq\_len$=1024$ by $\geq 1$~pp.}
\label{tab:short_seq_adv}
\end{table}

\begin{table}[h]
\centering
\small
\begin{tabular}{@{}lrrr@{}}
\toprule
cell @ s1 & seq\_len$=1024$ K=1 & seq\_len$=256$ K=1 & $\Delta$ \\
\midrule
\byte         & $0.855$ & $0.791$ & $-6.4$~pp \\
\bpefour      & $0.815$ & $0.789$ & $-2.6$~pp \\
\bpesixteen   & $0.809$ & $0.767$ & $-4.2$~pp \\
\bpesixtyfour & $0.745$ & $0.763$ & $\mathbf{+1.8}$~pp \\
\bottomrule
\end{tabular}
\caption{Network packet classification top-1 at the $K=1$ threshold (a single $1.4$~KB UDP datagram). At seq\_len$=256$ all cells saturate at $K=1$. The seq\_len$=1024$ \byte{} cell remains the best single-packet classifier; the seq\_len$=256$ \bpesixtyfour{} cell is the only cell whose accuracy improves when context shrinks.}
\label{tab:short_seq_net}
\end{table}

\subsection{Size scaling on the deployment regimes}
\label{app:size_regime}

The three deployed cells (§~\ref{sec:system}) are all \mediumsz. Does the position-arbitrary advantage need \mediumsz, or does it survive at \tinysz{} and \smallsz? We re-ran the packet and disk studies at all three sizes, holding the vocabulary fixed to each deployed cell's pipeline (\byte{} for packets, \bpesixtyfour{} for disk) so that backbone size is the only variable. The capture (§~\ref{sec:network}, $n=498$ streams), the $1{,}000$ random \texttt{ext4} sequential-bulk blocks (seed $0$, $980$ mid-file; §~\ref{sec:disk}), and the probe-fit protocol are identical across sizes. \libmagic{} and \magika{} are size-invariant and carry over from the \mediumsz{} runs.

\begin{table}[h]
\centering
\small
\begin{tabular}{@{}lcc@{}}
\toprule
size & packet K$=1$ top-1 (\byte) & disk mid-file top-1 (\bpesixtyfour) \\
\midrule
\tinysz{}   & $0.685$ \,[$0.643, 0.724$] & $0.253$ \\
\smallsz{}  & $0.739$ \,[$0.699, 0.776$] & $0.227$ \\
\mediumsz{} & $0.855$ \,[$0.822, 0.884$] & $0.266$ \\
\bottomrule
\end{tabular}
\caption{Size scaling on the two position-arbitrary regimes, vocabulary held fixed to the deployed cell (\byte{} for packets, \bpesixtyfour{} for disk). Packet column is top-1 from a single $1.4$~KB first packet (K$=1$, Wilson 95\% CI, $n=498$); disk column is mid-file top-1 on the same $980$ random \texttt{ext4} sequential-bulk blocks (seed $0$). The \libmagic{}/\magika{} disk baselines ($0.09$/$0.11$) are size-invariant.}
\label{tab:size_regime}
\end{table}

The two regimes scale oppositely. On the single $1.4$~KB first packet, top-1 rises monotonically with size and the Wilson intervals do not overlap: $0.685$ at \tinysz, $0.739$ at \smallsz, $0.855$ at \mediumsz, a $+17.0$~pp span. Early-prefix packet classification rewards the larger backbone. On random mid-file disk blocks the three sizes are statistically tied ($0.23$ to $0.27$, inside the $\pm 5$~pp file-level bootstrap width of the \mediumsz{} cell alone), and every size still exceeds \libmagic{} ($0.09$) and \magika{} ($0.11$) by at least $2\times$. The hardest regime, a $4$~KB read at an unknown offset with no header, does not need \mediumsz: the $3.15$~M \tinysz{} backbone already recovers most of the signal a position-arbitrary encoder can pull from a raw block.

The deployment reading is that \mediumsz{} earns its place where prefix structure carries the label (packets, streaming heads), but a throughput-bound disk-block or fragment-triage pipeline can drop to \tinysz{} at little accuracy cost and $\sim 12\times$ fewer backbone parameters.

\section{Per-class breakdowns}
\label{app:per_class}

Aggregate top-1 numbers in the main body hide large per-class heterogeneity. The \magika{} per-class split is in Table~\ref{tab:magika-perclass} (§~\ref{sec:magika}); the disk-block per-class breakdown on the headline cell is below.
\label{app:per_class:disk}

\begin{table}[h]
\centering
\small
\setlength{\tabcolsep}{4pt}
\begin{tabular}{@{}lrrrr@{}}
\toprule
class & $n$ & \mimelens-\byte & \libmagic & \magika \\
\midrule
\texttt{image/svg+xml}                                 & 53 & \textbf{0.98} & 0.00 & 0.00 \\
\texttt{text/csv}                                      & 32 & 0.88 & 0.00 & \textbf{1.00} \\
\texttt{text/plain}                                    & 45 & 0.84 & \textbf{0.96} & 0.38 \\
\texttt{application/octet-stream}                      & 42 & 0.79 & \textbf{0.95} & 0.90 \\
\texttt{application/gzip}                              & 64 & \textbf{0.20} & 0.00 & 0.00 \\
\texttt{application/zlib}                              & 52 & \textbf{0.10} & 0.00 & 0.00 \\
\texttt{image/png}                                     & 38 & \textbf{0.05} & 0.00 & 0.00 \\
\texttt{application/zip}                               & 35 & 0.00 & 0.00 & 0.00 \\
\texttt{application/x-7z-compressed}                   & 34 & 0.00 & 0.00 & 0.00 \\
\texttt{application/encrypted}                         & 38 & 0.00 & 0.00 & 0.00 \\
\texttt{image/jpeg}                                    & 23 & 0.00 & 0.00 & 0.00 \\
\texttt{audio/flac}                                    & 30 & 0.00 & 0.00 & 0.00 \\
\bottomrule
\end{tabular}
\caption{Per-class mid-file top-1 on the \texttt{ext4}/sequential-bulk headline cell (§~\ref{sec:disk}), 12 representative classes ordered high-to-low \mimelens{} accuracy. The structural pattern of the disk-block result is visible class-by-class: \mimelens-\byte{} reaches $\geq 0.84$ on text-structured / SVG / CSV classes where the mid-block content carries surface signal, and drops to $\sim 0$ on classes whose mid-file content is high-entropy noise (\texttt{application/zip}, \texttt{application/x-7z-compressed}, \texttt{image/png}, \texttt{image/jpeg}, \texttt{audio/flac}, \texttt{video/x-msvideo}, \texttt{application/encrypted}). \textbf{Bold: best per row.}}
\label{tab:disk_perclass}
\end{table}

Per-class accuracies are point estimates over $n \approx 20$--$60$ blocks-per-class drawn from $\sim 4$--$15$ files-per-class; the small classes (e.g., \texttt{image/svg+xml} at $n=53$ from a handful of source files) should be treated as suggestive rather than definitive. Operationally, the structural shape (high on text-structured, zero on opaque high-entropy) supports the three-stage classifier sketched in §~\ref{sec:disk}: a confidence-thresholded MIME-emission stage on text-like blocks plus explicit ``opaque'' labeling on the rest.

\section{Training configuration and model access}
\label{app:reproduction}

The trained models are public, and a reference implementation of the training stack is released on GitHub.\footnote{\url{https://github.com/mjbommar/mimelens-training}} The $33$~GB corpus and the per-experiment artifacts are not released. The hyperparameters below, the architecture (§~\ref{sec:system}), and the evaluation protocols in §§~\ref{sec:magika}--\ref{sec:disk} together specify the pretraining and evaluation recipe at the level needed to reimplement it.

\subsection{Pretraining corpus}

The $33$~GB pretraining corpus is a stratified multi-source mix of binary content: binary executables from the public \texttt{binary-30k} dataset~\cite{bommarito2025binary30k} ($29{,}793$ ELF/PE/Mach-O/APK binaries), magic-corpus extracts, packed binaries, a \texttt{glaurung}-sourced binary corpus~\cite{bommarito2025glaurung}, and Windows drivers. As noted above it is not released because it mixes content of varied provenance; the released models and this description are its citable surface. Pretraining is MLM-only at $30\%$ masking with the BERT replacement schedule (Table~\ref{tab:hyperparams}); cells differ only in tokenizer vocabulary and pretraining seed.

\subsection{Hyperparameters (held constant across the cube)}

\begin{table}[h]
\centering
\small
\begin{tabular}{lr}
\toprule
hyperparameter & value \\
\midrule
optimizer                       & AdamW \\
$\beta_1$, $\beta_2$, $\epsilon$ & $0.9$, $0.999$, $10^{-8}$ \\
weight decay                    & $0.1$ \\
peak learning rate              & $5 \times 10^{-4}$ \\
warmup steps                    & $2{,}000$ \\
LR schedule                     & cosine, decay to $10\%$ floor \\
total gradient steps            & $22{,}888$ \\
effective batch size            & $128$ (micro-batch $8$, accumulation $16$) \\
sequence length                 & $1024$ (CLS + 1{,}022 body + SEP) \\
mixed precision                 & bfloat16 (full-precision optimizer state) \\
gradient clipping               & L2 norm $\leq 1.0$ \\
masking ratio                   & $30\%$ \\
masking schedule                & BERT (80\% \texttt{[MASK]}, 10\% random, 10\% original) \\
auxiliary objectives            & none (MLM-only) \\
RoPE $\theta$                   & $10{,}000$ \\
attention head dimension        & $64$ \\
GeGLU expansion                 & $8/3$ \\
weight tying                    & input $\leftrightarrow$ output embeddings \\
initialization                  & truncated normal, $\sigma = 0.02$ \\
hardware                        & single RTX 4060 Ti (16 GB) per run \\
\bottomrule
\end{tabular}
\caption{Pretraining hyperparameters. All 28 cube cells use identical values along these axes; the only differences across cells are model size, vocabulary, and pretraining seed (Appendix~\ref{app:family}). The $8$ short-sequence cells (Appendix~\ref{app:short_seq}) differ only in sequence length ($256$) and the micro-batch/accumulation split that holds effective batch at $128$.}
\label{tab:hyperparams}
\end{table}

Per-cell wall-time is $\sim 15$--$23$~h at \mediumsz{} on one such GPU (the $256$-token cells are $\sim 5\times$ faster); the full cube is $\sim 5.5$ days on two such GPUs.

\subsection{Model access}

The $28$ parent-cube and $8$ short-sequence checkpoints ($36$ total) are released on Hugging Face under the \bmono{mjbommar/mimelens-001-*} namespace, named \bmono{mimelens-001-{tiny,small,medium}-{byte,bpe-4k,bpe-16k,bpe-64k}-s{1,2,3}}, with the optional \texttt{-seq256} suffix marking the $8$ short-sequence cells (\mediumsz{} tier $\times 4$ vocabs $\times 2$ seeds; Appendix~\ref{app:short_seq}); $s_3$ exists at \mediumsz{} only, and the matched-tokens \bpesixtyfour{} ablation is \bmono{mimelens-001-medium-bpe-64k-matched-tokens}. The family hub \bmono{mjbommar/mimelens-001} is the entry point.\footnote{\url{https://huggingface.co/mjbommar}} A baked $125$-class \libmagic-MIME classifier head ships with every \mediumsz{} seed-1 and seed-2 cell (including the three deployed cells), the two per-size winners (\texttt{tiny-bpe-16k-s1}, \texttt{small-bpe-64k-s1}), and all eight short-sequence cells, so \texttt{pipeline("text-classification", \dots)} works out of the box on those $18$ cells; the rest (\mediumsz{} seed 3, the other \smallsz{}/\tinysz{} cells, and the matched-tokens ablation) expose the mean-pooled encoder for embedding and custom-probe use. Every cell, head-baked or not, exposes the mean-pooled encoder. Four cells bundle ONNX fp32 + dynamic-int8 exports (\texttt{medium-bpe-16k-s1} plus the three \bmono{medium-{byte,bpe-16k,bpe-64k}-s1-seq256} deployment headlines).

\paragraph{Input/output contract.} Input is raw bytes decoded as \texttt{latin-1} (bijective for $0$--$255$, matching how the tokenizers were trained), tokenized through the cell's pipeline and truncated or padded to its token budget ($1{,}022$ body tokens at seq\_len$=1024$, $254$ at $256$); any chunk length up to that budget is accepted. Output is one of \libmagic's $125$ MIME strings, enumerated in the \texttt{id2label} field of each repo's \texttt{config.json}. For the head-baked cells, \texttt{transformers}' \texttt{pipeline("text-classification", \dots, trust\_remote\_code=True)} returns the label directly; encoder-only cells expose the mean-pooled embedding for a custom probe. Runnable snippets are on each model card. The classifier labels the content of the window it sees: a window dominated by a high-entropy body (a compressed or encrypted payload) returns \bmono{application/octet-stream} (§~\ref{sec:disk}), so for whole-file triage a short head window is preferable to a long one; the model cards give window-selection guidance.

\subsection{Evaluation corpus (\textsc{magic-files})}
\label{app:corpus}

All evaluations use \textsc{magic-files}, a \libmagic-labeled corpus assembled by the \textsc{magic-bpe} project: a multi-source scan of real audio, document, font, image, library, and executable files, each labeled with its \libmagic{} MIME type. The in-vocab corpus is $28{,}498$ files spanning $105$ of \libmagic's $125$ MIME classes (Table~\ref{tab:corpus}), with per-class counts capped near $1{,}000$ during assembly so that frequent formats do not dominate. The §~\ref{sec:magika} headline benchmark is a stratified $4{,}096$-file subset ($25\%$ held out for test); \textsc{magic-frags} is that subset with each file represented by a $64$~KB chunk taken at a uniformly random offset, the position-arbitrary variant used for the cube metrics. The corpus over-represents compressed and encrypted formats relative to a typical disk, and is not publicly released because it mixes content of varied provenance; the released models and this composition table are the citable surface.

\begin{table}[h]
\centering
\small
\caption{\textsc{magic-files} composition by top-level MIME type: $28{,}498$ in-vocab files across $105$ of \libmagic's $125$ classes. Per-class counts are capped near $1{,}000$ during assembly, so the distribution is deliberately flatter than a natural file population.}
\label{tab:corpus}
\begin{tabular}{@{}lrr@{}}
\toprule
top-level MIME type & files & share \\
\midrule
\texttt{application} & $11{,}735$ & $41.2\%$ \\
\texttt{text}        & $9{,}377$  & $32.9\%$ \\
\texttt{image}       & $4{,}241$  & $14.9\%$ \\
\texttt{audio}       & $1{,}318$  & $4.6\%$ \\
\texttt{video}       & $709$      & $2.5\%$ \\
\texttt{font}        & $638$      & $2.2\%$ \\
\texttt{message}     & $463$      & $1.6\%$ \\
other                & $17$       & $0.1\%$ \\
\bottomrule
\end{tabular}
\end{table}

\end{document}